\documentclass[aps,prx,reprint,superscriptaddress,nobibnotes,longbibliography]{revtex4-2} 
\pdfoutput=1

\usepackage{graphicx}
\usepackage{amsmath,amssymb}
\usepackage{xspace}
\usepackage{siunitx}
\usepackage[colorlinks]{hyperref}
\usepackage[caption=false]{subfig}
\usepackage{textcomp}
\usepackage{tikz,standalone}
\usepackage{float}

\newcommand{\ket}[1]{\ensuremath{\lvert #1 \rangle}\xspace}%
\newcommand{\avg}[1]{\ensuremath{\langle #1 \rangle}\xspace}%

\begin{document}

\title{Site-resolved observables in the doped spin-imbalanced triangular Hubbard model}

\author{Davis Garwood}
\author{Jirayu Mongkolkiattichai}
\author{Liyu Liu}
\author{Jin Yang}
\email[Corresponding author: ]{jy9ug@virginia.edu}
\author{Peter Schauss}
\email[Corresponding author: ]{ps@virginia.edu}
\affiliation{Department of Physics, University of Virginia, Charlottesville, Virginia 22904, USA}



\date{\today}

\begin{abstract}
The suppression of antiferromagnetic ordering in geometrically frustrated Hubbard models leads to a variety of exotic quantum phases including quantum spin liquids and chiral states.
Here, we focus on the Hubbard model on one of the simplest frustrated lattice geometries, a triangular lattice. Motivated by the recent realization of ultracold fermionic atoms in triangular optical lattices, we study the properties of the triangular-lattice Hubbard model through a Numerical Linked-Cluster Expansion algorithm. We investigate the Mott insulator transition finding a critical interaction $U_c/t = 7.0(2)$ and use spatial two- and three-point correlation functions to explore doped and imbalanced systems. Our results demonstrate that many interesting features occur at temperatures previously obtained for ultracold fermions in optical lattices and are accessible by upcoming experiments. Our calculations will be helpful for thermometry in ultracold atom quantum simulators and can guide experimental searches for exotic quantum phases in atomic triangular Hubbard quantum simulators. 
\end{abstract}

\maketitle

\section{Introduction}

Geometric frustration prevents a unique ground state in quantum many-body systems and thereby leads to a rich ground state phase diagram. One of the paradigm models of frustration are fermions with antiferromagnetic interactions on triangular lattice structures which were predicted to host exotic resonating valence bond states \cite{Anderson1987}.
In the focus of current research are quantum spin liquids \cite{Zhou2017}, which show no conventional long-range magnetic ordering down to zero temperature and can possibly be observed in a triangular Hubbard system \cite{Szasz2020}. Recent progress in numerical techniques and computing technology has revived the study of such frustrated systems \cite{Stoudenmire2011}.
In condensed matter experiments, signatures for such spin liquids have been recently found in materials with triangular and kagome lattice structure \cite{Zhou2017}. The investigation of frustrated structures has been pushed forward using spin systems realized by Rydberg atoms in optical tweezers \cite{Semeghini2021}. Such states have not been realized in Hubbard quantum simulators \cite {Gross2017} and the ground state properties of the triangular Hubbard model remain controversial due to limitations of the accessible system sizes in numerical calculations \cite{Yoshioka2009,Shirakawa2017,Szasz2020,Wietek2021,Chen2021,Cookmeyer2021}. 

Recently, quantum gas microscopy of bosons \cite{Yamamoto2020} and fermions \cite{Yang2021} in triangular-optical lattices has been demonstrated, clearing the path for the experimental investigation of  triangular-lattice atomic Hubbard models on the single-site and single-atom level. These microscopes are suited to detecting non-trivial ordering via spatial correlation functions. Motivated by this recent progress in quantum simulation of triangular Hubbard models, we investigate this system in the temperature regime relevant to ultracold atom experiments using Numerical Linked-Cluster Expansion (NLCE) \cite{Rigol2006,Rigol2007a,Rigol2007b,Khatami2011,Tang2012}. In particular, we focus on the study of spatial correlations which are directly accessible in quantum gas microscope experiments and verify our method using Determinantal Quantum Monte Carlo (DQMC) results based on the QuestQMC package \cite{Varney2009}. 

The Hubbard Hamiltonian on a two-dimensional triangular lattice with equal tunneling along all three lattice directions is given by \begin{equation} 
\label{eqn:1}
\begin{split}   
\mathcal{H} = -t\sum_{\avg{ij}} (\hat{c}_{i,\sigma}^\dagger \hat{c}_{j,\sigma} + \hat{c}_{j,\sigma}^\dagger \hat{c}_{i,\sigma}) + U \sum_i \hat{n}_{i,\uparrow} \hat{n}_{i,\downarrow}\\ -\sum_i [\mu (\hat{n}_{i,\uparrow} + \hat{n}_{i,\downarrow}) + h (\hat{n}_{i,\uparrow} - \hat{n}_{i,\downarrow})]\end{split}
\end{equation} 
where $\hat{c}_{i,\sigma}^\dagger(\hat{c}_{i,\sigma})$ is the creation (annihilation) operator for a fermion with spin $\sigma$ on site $i$, and $\hat{n}_{i,\sigma}=\hat{c}_{i,\sigma}^{\dagger}\hat{c}_{i,\sigma}$ is the particle number operator.
The first term describes the kinetic energy given by the creation and annihilation operators to all  sites with tunneling parameter $t$. $U$ is the on-site interaction and determines the interaction energy when a spin-up and spin-down fermion occupy the same site. These two terms are the core of the model and allow for the following occupancies at each site: no fermion, a spin-up fermion, a spin-down fermion, or two fermions with opposite spins \cite{Esslinger2010}. The parameters $\mu$ and $h$ together control the total particle density and spin imbalance. Importantly, these last two terms commute with the first two terms allowing the calculation of properties for any combination of $T$, $\mu$, and $h$ from a single exact diagonalization of $\mathcal{H}$ \cite{Khatami2011}.

The basis size of the Hubbard Hamiltonian scales as $4^N$ with the number of sites $N$ and, therefore, exact diagonalization of lattices is typically limited to 20 sites for ground states. For such lattices, finite size effects are still large and other techniques need to be employed. Though calculations based on matrix-product states have recently allowed the calculation of ground state properties of larger lattices \cite{Szasz2020}. Alternatively, DQMC techniques are challenging due to severe sign problems in the triangular Hubbard model \cite{Iglovikov2015,Mondaini2022}, and high-temperature series expansions are often limited to temperatures $T/t \gtrsim 1$. This is why we employ NLCE \cite{Rigol2006,Rigol2007a,Rigol2007b,Khatami2011,Tang2012} which enhances the capabilities of traditional series expansions at the cost of large numerical efforts of exact diagonalization of many clusters embedded in an infinite lattice. Our approach relies on the Hubbard Hamiltonian without additional approximations and the NLCE expansion provides the thermodynamic limit of the expectation value of any measurement operator.

\section{Numerical Linked-Cluster Expansion Implementation}

NLCE allow for the calculation of any extensive property $P$, a quantity dependent on system size, in the thermodynamic limit by summing over the weighted contributions of every cluster up to order $m$. 
\begin{align} 
&P(\mathcal{L})/N = \sum_{n=1}^{m}S_n, \\
&S_n = \sum_{c_n} L(c_n) \times W_P(c_n), 
\end{align} 
where $L(c_n)$ is the number of times a cluster of order $n$, $c_n$, can be embedded in lattice $\mathcal{L}$. The weighted property of $c_n$, $W_P(c_n)$, is given by the equation 
\begin{equation}
\label{eqn:weight}
W_P(c)=P(c)-\sum_{s \subset c} L_{c}(s) \times W_P(s),
\end{equation} 
here $s$ denotes the subclusters of cluster $c$ and $L_{c}(s)$ gives the number of times $s$ can be embedded in $c$. Due to this formulation, one only needs to consider connected clusters which can be naturally generated by progressively adding a chosen building block (site, triangle, square, etc.) depending on the periodic lattice geometry \cite{Rigol2006,Rigol2007a,Rigol2007b,Khatami2011,Tang2012}. Though in general it is easiest to expand by sites.

When diagonalizing the Hubbard Hamiltonian of a cluster we exploit atom number conservation to diagonalize the various atom number sectors separately. This reduces the memory required allowing higher orders to be reached on standard computer nodes. From the eigenenergies and eigenvectors of a subsector, the expectation value of any chosen extensive property can be calculated. These measurements are then thermally averaged using the sector's partition function and recombined to obtain the desired property of the cluster, $P(c)$. This general approach is easily applicable to other lattice Hamiltonians and can be adapted to different lattice structures.

\subsection{Expansion by sites}

Starting from a single site, in every step one site is added to an available edge of a cluster which creates a cluster of the next order. This process is repeated for all edges of all clusters of the same size. Since there are six edges per site for a triangular lattice, the number of clusters grows rapidly with the number of sites. However, many of these clusters are translations of each other and, due to the formulation of NLCE, can be discarded reducing the overall cluster count \cite{Rigol2006,Rigol2007a,Rigol2007b,Khatami2011,Tang2012}. If a different expansion scheme is used, the convergence behavior of a property, especially at low temperature, can differ.

\begin{figure}[h]
    \centering
    \includegraphics{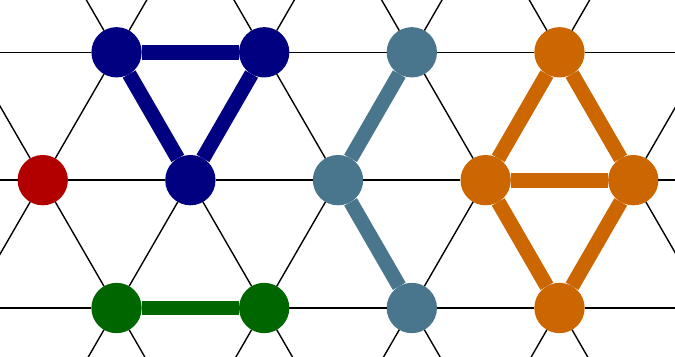}
    \label{fig:siteExpClusters}
    
\vspace{0.6cm}
\begin{tabular}{m{0.8cm}|m{1.5cm}|m{1.4cm}||m{0.8cm}|m{1.5cm}|m{1.4cm}}
Order & Connected Clusters& Top.\,Dist. Clusters & Order & Connected Clusters & Top.\,Dist. Clusters \\\hline
1 & 1 & 1 & 6 & 814 & 22\\
2 & 3 & 1 & 7 & 3652 & 54\\
3 & 11 & 2 & 8 & 16689 & 156\\
4 & 44 & 4 & 9 & 77359 & 457\\
5 & 186 & 8 & 10 & 362671 & 1424
\end{tabular} 
    \caption{{\bfseries Triangular Clusters} Top, first few clusters obtained through site expansion on an infinite triangular lattice. Bottom, number of connected and topologically distinct clusters on this lattice for orders (number of sites) up to order 10 of an expansion by sites \cite{Rigol2007a}.}
    \label{tab:Table1}
\end{figure}

As shown in Fig.~\ref{tab:Table1}, the number of connected clusters remains significantly large for higher orders, increasing the calculational effort tremendously. For this reason, it can be useful to consider alternative expansion schemes such as an expansion by triangles (see Appendix B) which require fewer clusters in the higher orders. 

\subsection{Reduction of cluster number \& identifying subclusters}

Since exact diagonalization and thermal averages are time-consuming and have to be performed for every cluster in the expansion, a critical optimization is reducing the number of required clusters in the expansion. Isomorphic clusters, that is clusters with the same graphical structure, have the same Hubbard Hamiltonian. The multiplicity, $L(c)$, of a unique cluster, $c$, is then given by the number of isomorphisms \cite{Rigol2006,Rigol2007a,Rigol2007b,Khatami2011,Tang2012}. Another approach is to first identify symmetric clusters. However, we found it easier and more efficient to leverage optimized isomorphism algorithms to reduce directly to topologically distinct clusters. The downside to this approach is that the proper cluster weighting needed for long-range properties beyond nearest-neighbor is not preserved.

The subcluster multiplicities of every cluster that are required to apply Eq.~\ref{eqn:weight} can be found in a similar manner. Keeping track of all the unique clusters that lead to the current cluster during the expansion phase, we iterate over this list counting the number of subgraph isomorphisms $L_c(s)$ for each subcluster. With this information, the Hamiltonians of all clusters up to order $m$ can be constructed from their adjacency matrices and then diagonalized. 

\subsection{Resummation}

Once the orders of a property have been determined, a resummation algorithm is used to improve convergence at lower temperatures. We found that Wynn's epsilon method led to the best results, though other resummations, such as the Euler transformation, can be suitable for particular properties \cite{Rigol2007a, Tang2012}. The error on such a resummation, can be estimated by the difference between subsequent orders or the difference between different resummation techniques.

The Wynn resummation algorithm is defined by the equation
\begin{equation} 
\epsilon_n^{(k)}=\epsilon_{n+1}^{(k-2)}+ \frac{1}{\epsilon_{n+1}^{(k-1)}-\epsilon_n^{(k-1)}},
\end{equation} 
with base cases,
\begin{equation}
    \epsilon_n^{(-1)}=0, ~~ \epsilon_n^{(0)}=P_n.
\end{equation} 
Only the even terms $\epsilon_n^{2l}$, where $l$ is the number of cycles of improvement, are considered as they are the only terms expected to converge. 

Euler's transformation is specific to alternating series. The partial sums $S_n$ become $u_n= (-1)^nS_n$ such that the summation is
\begin{equation}
P(\mathcal{L})/N= u_0-u_1+...-u_{n-1}+\sum_{l=0}^{m-n}\frac{(-1)^l}{2^{l+1}}\Delta^l u_n.
\end{equation} 
Here $\Delta$ denotes the forward difference operator 
\begin{equation}
\Delta^l u_n = \Delta^{l-1}u_{n+1} - \Delta^{l-1}u_n,
\end{equation} 
and $n-1$ is the term after which Euler's transformation is applied \cite{Rigol2007a, Tang2012}. The absolute difference between the maximum order Wynn resummation and maximum order Euler resummation is used as the default error estimate.

\section{NLCE of Triangular-lattice Hubbard Systems}

Here we apply the NLCE expansion approach to the triangular-lattice Hubbard model without any additional approximations. At small $U$ and at large $U$ approximations allow a detailed study of the triangular-lattice Hubbard model. Therefore, we focus on intermediate interactions where the situation is less clear. Using NLCE we can reach similar temperatures as the DQMC calculations and in particular, we can cover temperatures $T/t \sim 0.3 \ldots 1$ close to the Mott regime which is the most relevant regime for current ultracold atom experiments. In this temperature range, we focused on properties corresponding to experimentally measurable quantities in quantum gas microscopes \cite{Endres2013}. Starting with a comparison to DQMC calculation, we discuss the Mott insulator (MI) transition in the triangular Hubbard system, and study various spatial correlation functions, including spin-density correlations and three-point correlations. All of these correlations are accessible with currently available quantum gas microscopy technology \cite{Koepsell2020,Hartke2020}.

\subsection{Comparison to Determinantal Quantum Monte Carlo}

To cross-check the accuracy of our NLCE implementation we performed DQMC calculations. DQMC is an efficient technique to simulate the properties of fermions in lattices \cite{Blankenbecler1981,Paiva2010}.
For the DQMC calculations, we rely on a Fortran 90/95 package, QUantum Electron Simulation Toolbox (QUEST)
\cite{Varney2009}.

\begin{figure}[H]
    \centering
	\includegraphics[width=1\columnwidth]{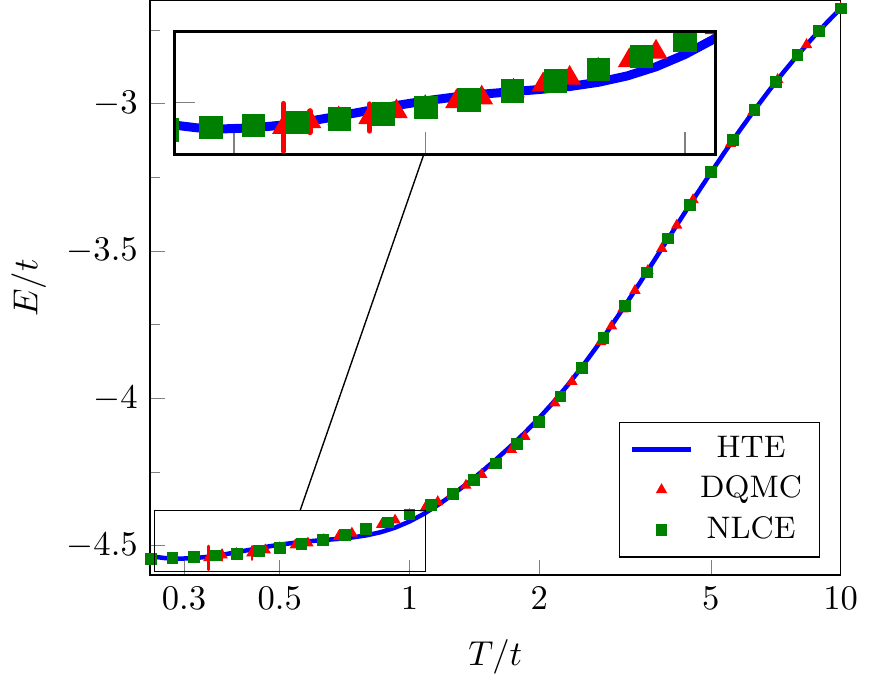}
	\caption{{\bfseries Energy per site in the triangular Hubbard model.} 9th order NLCE expansion by sites (with 1 cycle of Wynn resummation) of a spin-balanced triangular system at $U/t=8.0$ and $\mu/t=4$ as compared to DQMC and a high temperature series expansion (HTE) \cite{Henderson1992}. The HTE curve shows small deviations at temperatures $T/t \lesssim 1$. DQMC results  for $T/t \lesssim 0.3$ are unreliable due to a severe sign problem. DQMC and NLCE agree within error bars over the complete range. The NLCE converges down to lower temperatures than the DQMC here. }
	\label{fig:E}
	
\end{figure}

For triangular lattices, the calculations suffer from a severe sign problem when approaching low temperatures \cite{Iglovikov2015} and we rely on extended averaging for low temperatures. Calculations are stopped when the sign is approximately zero within error bars. Although the sign problem is severe, reliable results were obtained down to similar temperatures as NLCE over wide parameter regimes, at a cost of dramatically increased computation times.

Simulations rely on a homogeneous 8$\times$8 lattice with periodic boundary conditions. We confirmed that for the properties discussed in this manuscript, finite-size errors are smaller than combined Trotter and statistical error. The inverse temperature $\beta=L d\tau$ was split into $L=200$ imaginary time slices. To obtain higher statistics, the simulations were averaged over ten or more runs, 50,000 passes each. This allows us to control imaginary time correlations and sampling errors by comparing the variance from each individual run with the variance of all runs.

The comparison of DQMC and NLCE calculations with very different error sources allows for more conclusive results at lower temperatures where each of the individual techniques alone becomes questionable. For short-range properties, NLCE was found to be simultaneously more manageable and faster when performing calculations for large parameter ranges. This is because NLCE, as previously mentioned, allows for calculation of properties at multiple temperatures, $\mu$ values, and $h$ values without rediagonalization, and the needed diagonalizations can be easily run in parallel. Conversely, a DQMC requires a separate run for every parameter combination. It is also worth noting that with NLCE, a low order scan of parameters and properties can be performed with little time and resources to narrow down regimes and measurements of interest without worry of the sign problem.

An illustrated comparison of these two techniques for the energy per site is depicted in Fig.~\ref{fig:E}. By going up to order 9 and then performing a cycle of Wynn's method, NLCE achieved lower temperatures with significantly lower error than DQMC for this property in less time and with less computing resources. We find very good agreement between our NLCE and the DQMC results. For temperatures $T \gtrsim 1$ the high-temperature series expansion is an excellent approximation. NLCE convergence using calculations up to order 9 is sufficient here to reach temperatures down to the lowest that have been reached in experiments with ultracold fermions in optical lattices of $T/t \sim 0.25$ \cite{Mazurenko2017}.

\subsection{Compressibility near Mott regime}

\begin{figure}[h]
    \centering
    \includegraphics[width=1\columnwidth]{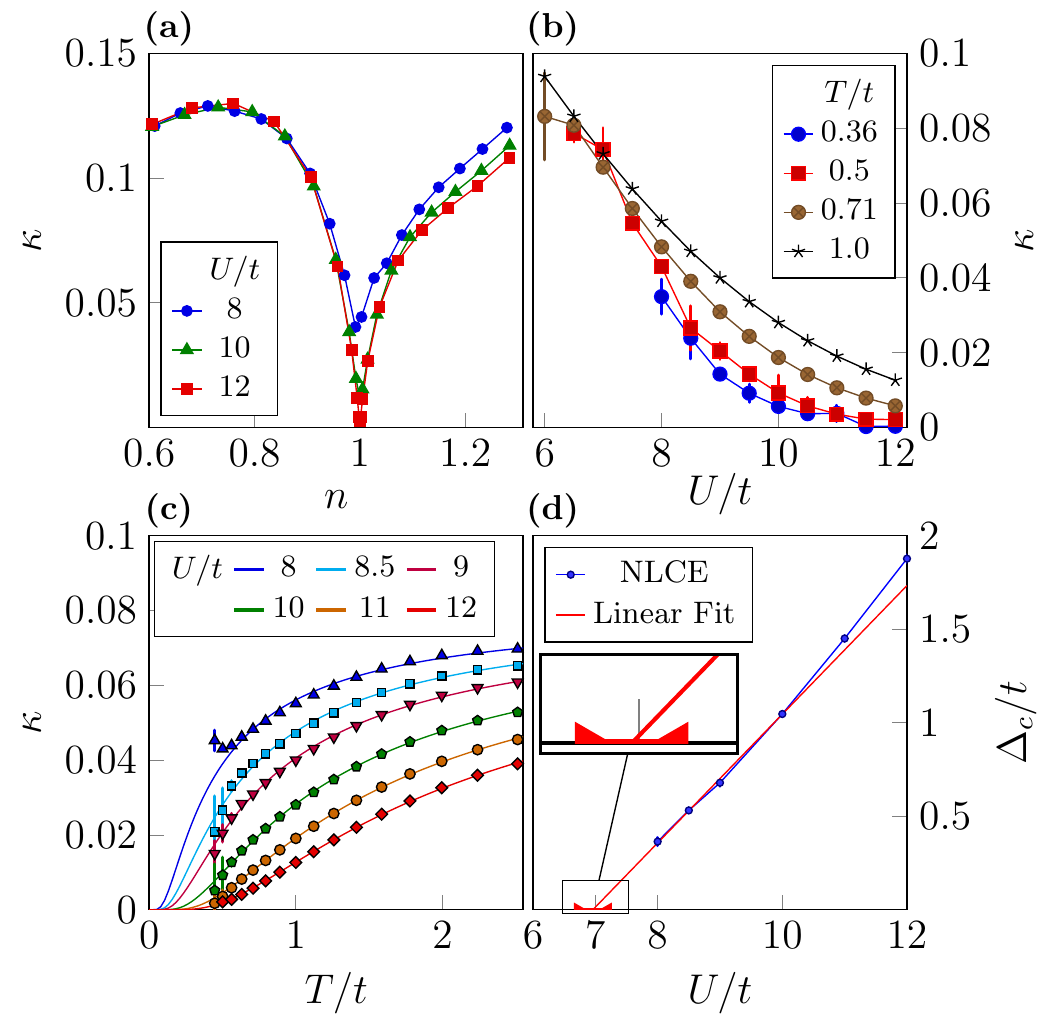}
    \caption{{\bfseries Compressibility and MI transition} {\bf(a)} Compressibility versus atom number density at $T/t=0.5$. {\bf(b)} Compressibility versus interaction $U/t$. Only data that is sufficiently convergent is shown. {\bf(c)} Compressibility versus temperature for interaction $U/t = 8$ to 12. Fits are $\kappa(T) = a \exp(-\Delta_c/T)$ \cite{Kokalj2013}. {\bf(d)} Scaling of $\Delta_c$ from (c) to determine the Mott transition. An error-weighted fit yields $U_c/t = \num{7.0(2)}$. The 4th order Wynn resummation of the compressibility was used (error with respect to 9th order Euler resummation).}
    \label{fig:compressibility}
\end{figure}

The MI transition is of particular interest because of its relation to high-temperature superconductivity \cite{Lee2006}. It occurs for strong interactions ($U/t \gg 1$) and low temperature ($k_{B}T \ll U$) where double occupancy is suppressed \cite{Gebhard1997}. The Mott insulating phase is characterized by a vanishing compressibility. We calculate the compressibility $\kappa = \frac{\partial n}{\partial\mu}$ over a wide parameter range at half filling (density $n=1$). Note, that the triangular-lattice Hubbard model does not show particle-hole symmetry requiring to find the correct chemical potential $\mu$ for each temperature to obtain exactly $n=1$ (see Appendix D).

The sharp drop in compressibility shown in Fig.~\ref{fig:compressibility} is indicative of the MI transition and allows us to approximate the region in which the lattice is in this insulating phase. By calculating the compressibility versus interaction $U/t$ we find a sharpening transition upon lowering the temperature (see Fig.~\ref{fig:compressibility}(b)). To determine the approximate value of $U/t$ for the MI transition in the triangular Hubbard model, we perform an extrapolation using a gap model \cite{Kokalj2013}. We find a linear closing of the gap $\Delta_c$ and a zero temperature transition point $U_c/t = \num{7.0(2)}$ (see Fig.~\ref{fig:compressibility}(d)). This is consistent with $U_c/t\sim 7$ found in refs.~\cite{Clay2008,Kokalj2013}, but shows a deviation from other results \cite{Kokalj2013supp,Aryanpour2006,Galanakis2009,Dang2015,Wietek2021}. We note that the Mott transition discussed here is at different $U/t$ than the transition to $\SI{120}{\degree}$ order which occurs at higher $U/t$. 

\subsection{Thermometry of triangular Hubbard systems}

Thermometry of ultracold atomic systems in optical lattices in the Hubbard regime is challenging. Previous experiments mostly relied on comparison of measured spin-spin correlation functions to calculations to determine the temperature of the system \cite{Parsons2016,Brown2017}. Here, we demonstrate that such an approach is also possible for a triangular-lattice Hubbard model although the presence of sufficiently strong spin correlation is less obvious due to the geometric frustration. In ultracold atom lattice experiments, the spin operator $S_z$ is defined by 

\begin{equation}
    \hat{S}_{z,i}= \frac{1}{2} (\hat{n}_{i,\uparrow}-\hat{n}_{i,\downarrow} ),
\end{equation} 
with $\hat{n}_{i,\sigma} = \hat{c}_{i,\sigma}^\dagger \hat{c}_{i,\sigma}$.
While the spin-spin correlations are isotropic at $h=0$, the $S_z$ basis is selected by the atom number measurement in the experiment. The spin-imbalance $h$ is typically introduced in the atom-number sector and therefore along the same dimension.
We consider nearest-neighbor two-point correlators given by the relation
\begin{equation}
    C_1(\boldsymbol{r_1},\boldsymbol{r_2}) = \avg{\hat{A}_1 \hat{B}_2}-\avg{\hat{A}_1}\avg{\hat{B}_2},
\end{equation} 
where $\hat{A_1}$ and $\hat{B_2}$ are arbitrary quantum operators at sites \textit{\textbf{r$_1$}} and \textit{\textbf{r$_2$}}.

\begin{figure}[h]
	\centering
	\includegraphics[width=1\columnwidth]{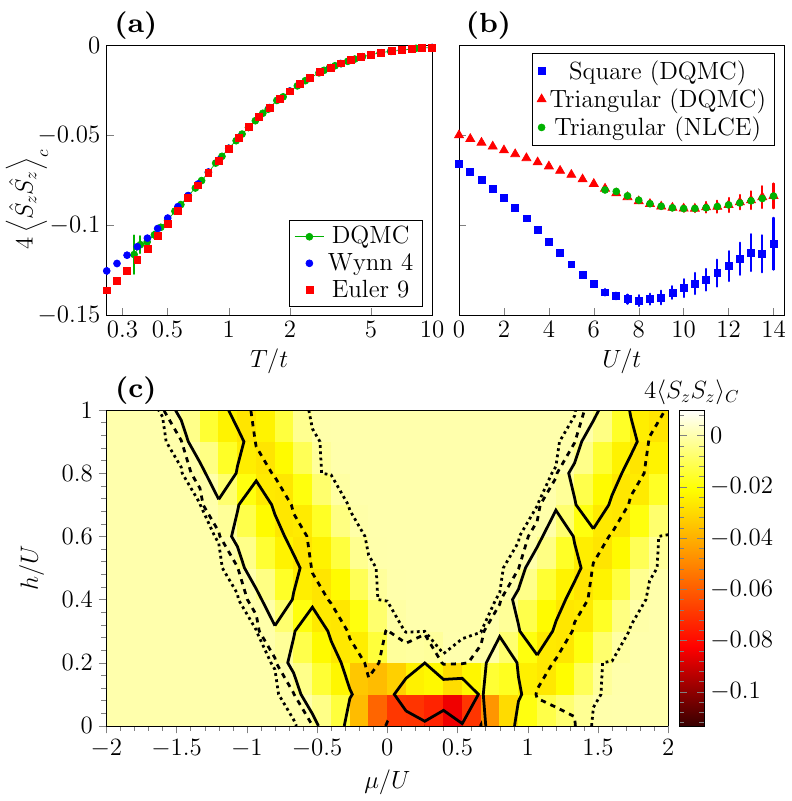}
	\caption{ {\bfseries Spin-spin correlations (a)} Spin-spin correlator versus temperature at $U/t=8$, $\mu/t=4$, and $h/t=0$. The 4th order Wynn resummation (Wynn 4) and the 9th order Euler resummation (Euler 9) do not agree at low temperatures and we use their difference to estimate the uncertainty. {\bf(b)} Spin-spin correlator versus interaction at $T/t=0.6$, $\mu=U/2$, and $h/t=0$. Nearest-neighbor spin-spin correlation are maximally anti-correlated near $U/t \sim 10$ for the triangular geometry. This is in contrast to the 2D square Hubbard model where a maximum occurs closer to $U/t \sim 8$ \cite{Cheuk2016}. NLCE does not converge well for $U/t \lesssim 6$. {\bf(c)} Spin-spin correlations in the $\mu/U$-$h/U$ plane at $U/t=8$ and $T/t=0.355$ with 4 cycles of Wynn resummation. Dotted, dashed, and line contours enclose areas with $10^{-4}$, $10^{-3}$, and $10^{-2}$ uncertainty (relative to 9th order Euler resummation), respectively.}
	\label{fig:sz_sz_tri}
\end{figure}

We calculate the $\hat{S_z}$-$\hat{S_z}$ spin correlation versus temperature $T$ at $U=8$ and find a dependence on temperature which is useful for thermometry. Our NLCE results match the DQMC results within error bars down to about $T/t \approx 0.4$ where the sign problems starts to make DQMC calculations intractable (see Fig.~\ref{fig:sz_sz_tri}(a)). 

The interaction dependence of antiferromagnetic correlations in a triangular lattice follows a similar behavior as in a square lattice but antiferromagnetic correlations become maximal around $U \approx 10$ for a triangular lattice  compared to $U \approx 8$ in the square lattice \cite{Parsons2016,Cheuk2016} (Fig.~\ref{fig:sz_sz_tri}). In addition, we find that the magnitude of correlations in a triangular lattice is only about half compared to the square lattice case at temperatures around $T/t = 0.6$. Within the range of NLCE calculations $U/t > 6$ we find very good agreement between DQMC and NLCE results.

Although the system is frustrated, spin-spin correlations are sufficient to provide a useful thermometer. We find generally antiferromagnetic correlations at nearest-neighbor sites and observe that all three pairs of spins on a triangular plaquette are anti-correlated. This is consistent with expectations for \SI{120}{\degree} ordering but we cannot make predictions about long-range ordering with our approach. 

\section{Spin-density correlations}

\begin{figure}[h]
    \centering
    \includegraphics[width=1\columnwidth]{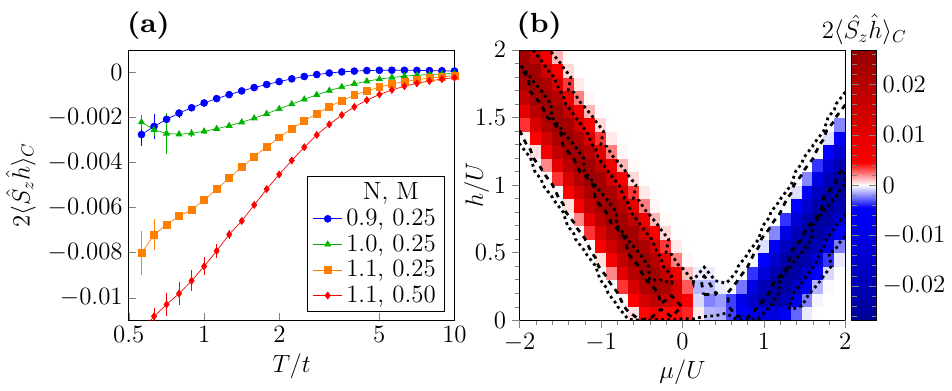}
    \caption{ {\bfseries Nearest-neighbor spin-hole correlator (a)} Versus temperature at $U/t=8$ with the doping, $N=n_\uparrow+n_\downarrow$, and spin imbalance, $M=n_\uparrow-n_\downarrow$, selected for. 9th order Euler resummation is plotted with error calculated in respect to the 4th order Wynn resummation. {\bf(b)} In the $\mu/U$-$h/U$ plane at $U/t=8$ and $T=0.5$ and with 4 cycles of Wynn resummation. Dotted and dashed contours enclose areas with $10^{-4}$ and $10^{-3}$ uncertainty (relative to 9th order Euler resummation), respectively. We note that at half filling ($\mu/U \approx 0.5$), the correlations are negative.}
    \label{fig:spin_hole}
\end{figure}

Spin and density typically decouple to a large degree in many Hubbard systems. In one-dimensional systems, the decoupling is perfect and leads to the phenomenon of spin-charge separation \cite{Vijayan2020}.
As an example of spin-charge coupling, we study the correlation $\avg{\hat{S}^z_i \hat{h}_j}_c = \avg{\hat{S^z}_i \hat{h}_j} - \avg{\hat{S}^z_i} \avg{\hat{h}_j}$ where $\hat{h}_j=1-\hat{n}_j$.
Due to spin-inversion symmetry, this correlator vanishes in spin-balanced systems. However, for doped spin-imbalanced, $\mu, h \neq$ 0, systems we find non-zero correlations even at densities $\geq1$ where few holes are expected (Fig.~\ref{fig:spin_hole}). 
Around half-filling ($\mu/U \approx 0.5$) we find negative correlations extending into the spin-imbalanced regime ($h > 0$).
We interpret this as signatures for repulsive pairing effects between the minority ($\ket{\downarrow}$) and holes as discussed in refs.~\cite{Zhang2018, Morera2021}. This spin-hole relationship can also be probed experimentally. Measurements require spin- and density-resolved imaging in a quantum gas microscope which has been demonstrated for square lattices already \cite{Koepsell2020,Hartke2020}. These experimental techniques are also applicable to triangular optical lattices.

\subsection{Three-point correlations}

\begin{figure}[h]
    \centering    
    \includegraphics[width=1\columnwidth]{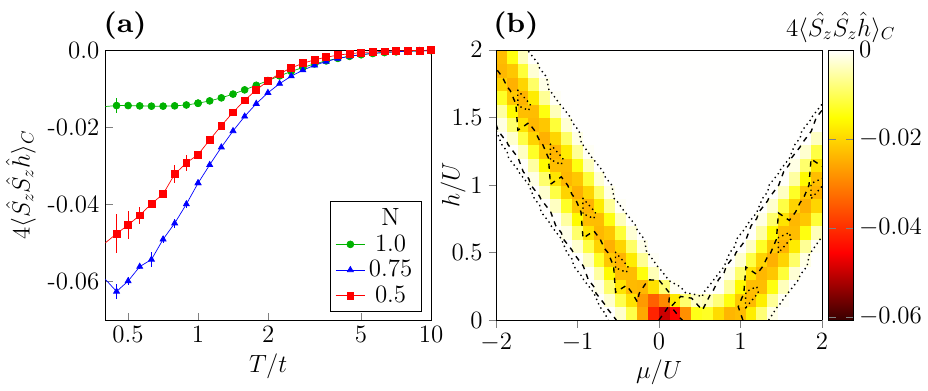}
    \caption{ {\bfseries Spin-spin-hole correlator on a triangular plaquette (a)} Versus temperature with 4 cycles of Wynn resummation for different dopings, $U/t=8$, and spin balanced. {\bf(b)} In the $\mu/U$-$h/U$ plane at $U/t=8$ and $T=0.5$ and with 4 cycles of Wynn resummation. Dotted and dashed contours correspond to $10^{-4}$ and $10^{-3}$ uncertainty (relative to 9th order Euler resummation), respectively.}
    \label{fig:spin_spin_hole}
\end{figure}

To demonstrate the power and flexibility of NLCE calculations, we go beyond two-point correlators and calculate three-point correlators. Those are naturally appearing in the triangular Hubbard system due to the three sites in each plaquette of the lattice. In general, a connected three-point correlator is given by 
\begin{equation}
    \begin{split}
    C_3(\boldsymbol{r_3},\boldsymbol{r_2},\boldsymbol{r_1}) = \avg{\hat{Z_3}\hat{Y_2}\hat{X_1}} - \avg{\hat{Z_3}}\avg{\hat{Y_2}\hat{X_1}} -\\ \avg{\hat{Y_2}}\avg{\hat{Z_3}\hat{X_1}} - \avg{\hat{X_1}}\avg{\hat{Z_3}\hat{Y_2}} + 2\avg{\hat{Z_3}}\avg{\hat{Y_2}}\avg{\hat{X_1}}
    \end{split}
\end{equation} 

where $\hat{X_1}$, $\hat{Y_2}$, and $\hat{Z_3}$ are arbitrary quantum operators at nearest-neighbor sites \textit{\textbf{r$_1$}}, \textit{\textbf{r$_2$}}, and \textit{\textbf{r$_3$}}, respectively \cite{Endres2013,Koepsell2021}. With this correlator we can examine the spin-spin-hole correlation on a triangle, $\langle \hat{S_z},\hat{S_z},\hat{h} \rangle_C$, which averages the three point correlators of the three orderings of $\hat{S_z}$, $\hat{S_z}$, and $\hat{h}$. In particular, this correlation measures the perturbation of the antiferromagnetic correlations due to the presence of a hole (Fig.~\ref{fig:spin_spin_hole}). In contrast to the square-lattice case \cite{Koepsell2021}, we find negative three-point correlations for all dopings and spin-imbalances studied.
We interpret this as enhancement of antiferromagnetic correlations due to the reduction of frustration caused by the absence of the third spin on a triangular plaquette.

\section{Conclusion and Outlook}

Using NLCE on a triangular lattice proved to be a powerful computational approach without sign problem, and will be a useful tool for comparison to future experimental data in quantum gas microscopes. In particular, we found it to perform favorably in the strong interaction regime $U/t \gtrsim 7$. We used it to study a variety of properties of the triangular Hubbard model in regimes that are accessible via quantum simulation in recently established ultracold atom experiments \cite{Yang2021}. To benchmark our NLCE algorithm, we compared NLCE calculations with DQMC simulations and found very good agreement in the regimes where both techniques converge. 

Using a large dataset for temperature-dependent compressibility we found the critical interaction for the Mott transition. Nearest-neighbor spin correlation functions were examined for both triangular and square Hubbard model. Though, there were similarities between their spin correlation functions the correlations in the triangular Hubbard model are suppressed compared to the square Hubbard model. Lastly, we investigated spin-density correlations and demonstrated the calculation of short-ranged multi-point correlations on triangular plaquettes using NLCE. It is possible to calculate beyond nearest-neighbor correlations, but we expect that higher orders of the triangular-lattice expansion would be necessary for reliable results which are probably reachable with more advanced or approximate diagonalization techniques \cite{Bhattaram2019}. Our calculations demonstrate the possibilities of quantum simulations of triangular Hubbard systems in experiments and many of the discussed features are within experimental reach. Future NLCE studies on triangular lattices may be able to give access to transport properties \cite{Vranic2020} and chiral ordering \cite{Wen1989,Szasz2020}. 

\raggedbottom
\begin{acknowledgments}
We acknowledge discussions with Bruno R. de Abreu, Eugene Demler, Ehsan Khatami, Israel Klich, Tommaso Macr\`i, Vinicius Z. Pedroso, and Rajiv Singh. We thank Gia-Wei Chern for comments on the manuscript and Jaan Oitmaa for sharing the high-temperature series expansion coefficients for the triangular lattice \cite{Henderson1992}.
We acknowledge support by the NSF CAREER award (\#2047275), the Thomas F. and Kate Miller Jeffress Memorial Trust and the Jefferson Trust.
D. G. was supported by a Ingrassia Scholarship.
J. M. acknowledges support by The Beitchman Award for Innovative Graduate Student Research in Physics in honor of Robert V. Coleman and Bascom S. Deaver, Jr.
\end{acknowledgments}

\bibliographystyle{apsrev4-2}

%

\section{Appendix}

\subsection{Alternative expansion by triangles}

Instead of expanding by sites, an infinite triangular lattice can also be constructed via an expansion using triangle. Only upwards or downwards facing triangles, in addition to a single site, are considered. This is because each edge is either a part of a unique upwards or downwards facing triangle to maintain consistency. The choice of orientation is arbitrary, but one must have a way of differentiating the triangle orientation. Note that the multiplicity, $L(c)$, should be weighted by $1/3$, for all $n>0$ \cite{Rigol2007a}. 

\vspace{0.6cm}
\begin{figure}[h]
    \centering
    \includegraphics{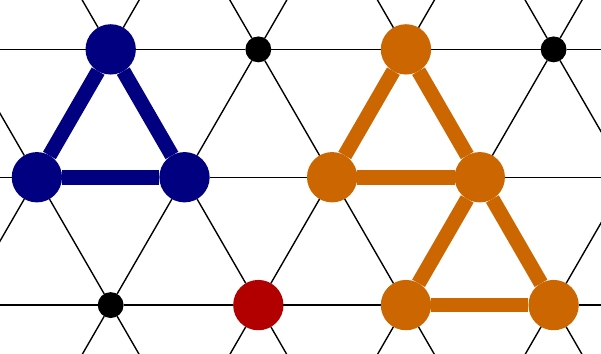}
\vspace{0.6cm}
\begin{tabular}{m{0.8cm}|m{1.5cm}|m{1.4cm}||m{0.8cm}|m{1.5cm}|m{1.4cm}}
Order & Connected Clusters & Top.\,Dist. Clusters & Order & Connected Clusters & Top.\,Dist. Clusters \\\hline
1 & 1 & 1 & 4 & 44 & 5\\
2 & 1 & 1 & 5 & 186 & 12\\
3 & 3 & 1 & 6 & 814 & 35
\end{tabular}
    \caption{{\bfseries Clusters via triangle expansion.} Top, the first few clusters obtained through a triangle expansion on an infinite triangular lattice. Bottom, the number of connected and topologically distinct clusters on this lattice for orders (number of triangles) up to order 6 \cite{Rigol2007a}.}
    \label{tab:Table2}
\end{figure}

The results obtained through an expansion by triangles (Fig.~\ref{tab:Table2}) converge to those of an expansion by sites at high temperatures. Though the convergence properties at low temperatures will vary. For specific properties, it was observed that the convergence at low temperatures was better for a triangular expansion than for a site expansion. On the other hand, when resumming the results, sharp discontinuities tend to be introduced more often. Consequently, we use the expansion by sites as the default expansion scheme as it was in general more reliable for most properties.

\subsection{Absence of particle-hole symmetry}

\begin{figure}[h]
    \centering
    \includegraphics[width=1\columnwidth]{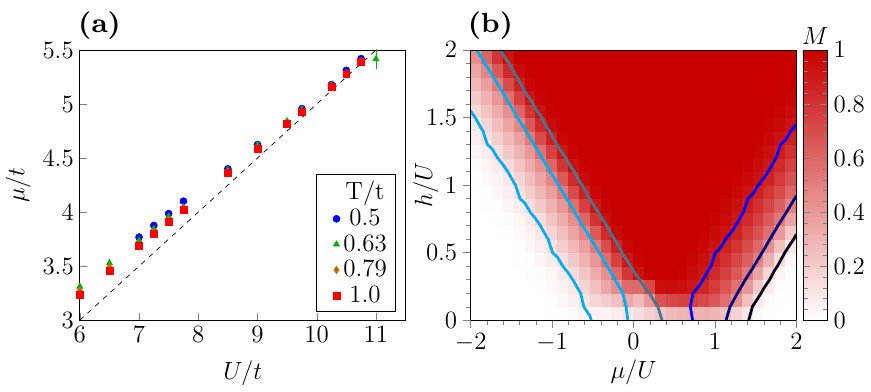}
    \caption{{\bfseries (a) Deviation of $\mu$ at half-filling from $U/2$.} For lower temperatures and smaller interactions there is a clear deviation from the $\mu=U/2$ line. We find that the deviation from particle-hole symmetry increases with lower interaction and lower temperature.{\bfseries (b) Spin imbalance and density in $\mu/U-h/U$ plane.} From left to right, the drawn contours represent 0.1, 0.5, 0.9, 1.1, 1.5, and 1.9 filling. $T/t= 0.5$ and $U/t= 8.0$.}    
    \label{fig:density_vs_mu_h}
\end{figure}

In contrast to Hubbard models with particle-hole symmetry, where half-filling (density $n=1$) is always obtained at $\mu=U/2$, for a Hubbard model on a triangular lattice the $\mu$ for half-filling has a more complicated dependence on the model parameters (Fig.~\ref{fig:density_vs_mu_h}{\bf(a)}). For any figure with specified density, the density data was interpolated to determine the $\mu$ value corresponding to half-filling. This value is then used with another interpolation of the property being examined to determine its value at half-filling. Additionally, when $h \neq 0$, density becomes interconnected with spin imbalance. This density-imbalance relationship is exemplary shown for temperature ($T/t=0.5$) in Fig.~\ref{fig:density_vs_mu_h}{\bf(b)}.

\subsection{Spin-hole correlations in square Hubbard model}

\begin{figure}[H]
    \centering
    \includegraphics[width=1\columnwidth]{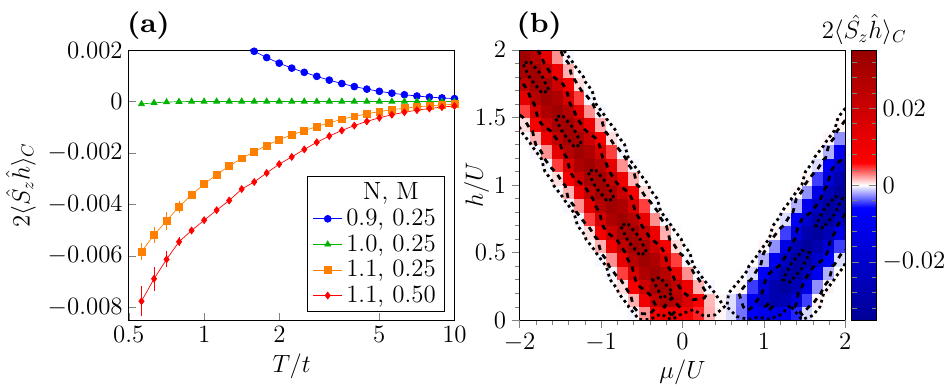}
    \caption{ {\bfseries  Spin-hole correlator on a square lattice (a)} Versus temperature at $U/t=8$ with the doping and spin imbalance selected for. 9th order Euler resummation is plotted (error with respect to 4th order Wynn resummation). {\bf(b)} 4th order Wynn resummation in the $\mu/U$-$h/U$ plane at $U/t=8$ and $T=0.5$. Dotted and dashed contours enclose areas with $10^{-4}$ and $10^{-3}$ uncertainty (relative to 9th order Euler resummation), respectively.}
    \label{fig:spin_hole_sqr}
\end{figure}

On a square lattice the spin-hole correlator goes to zero at half-filling (Fig.~\ref{fig:spin_hole_sqr}). This is in contrast to a triangular lattice where we instead see a negative correlator in the region around half-filling. Elsewhere, the correlators follow the same overall pattern of the same order of magnitude. Consequently, this region about half-filling would seem the most promising for looking for spin-hole pairing that emerges on a frustrated lattice.

\subsection{Extrapolation of doublon density to {\it T} = 0}
\begin{figure}[H]
    \centering
    \includegraphics[width=1\columnwidth]{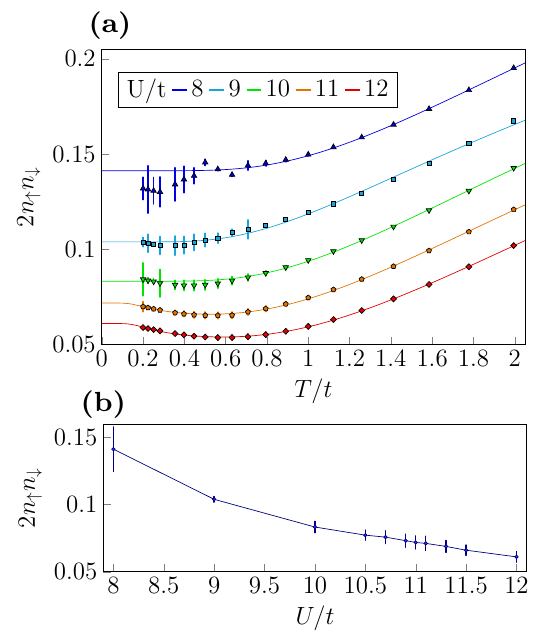}
    \caption{{\bfseries Doublon density. (a)} We extrapolate the doublon density to $T/t=0$ from a model fit using NLCE calculations for $U/t = 8\ldots12$ in the temperature range $T/t = 0.2\ldots 2.5$. We do not show all NLCE data to avoid overlapping points. {\bfseries (b) } We find a smooth dependence of the zero-temperature doublon density with $U/t$ within error bars. Convergence for $U/t < 8$ is not sufficient to explore the doublon density over the Mott insulator phase transition around $U/t \approx 7$.}    
    \label{fig:doublon_extrap}
\end{figure}

In previous studies, steps in the $T = 0$ doublon density versus interaction have been observed \cite{Yoshioka2009,Shirakawa2017}. Therefore, we extrapolate the doublon density calculated from NLCE to zero temperature (Fig.~\ref{fig:doublon_extrap}). We use an empiric scaling function that is inspired by a gap model: $a_1 e^{-b/T} - a_2 e^{-b c /T^2} + d$. Here, $a_1,a_2,b,d$ are fit variables for each $U$ while $c$ is a global fit variable. The variable $d$ yields the doublon fraction at zero temperature. We do not see steps when extrapolating the doublon density to zero temperature. The step size previously observed was rather small and may be below the resolution of our extrapolation.

\end{document}